\documentclass[twocolumn,,preprintnumbers,amsmath,amssymb]{revtex4}

\usepackage{graphicx}
\usepackage{dcolumn}
\usepackage{bm}
\usepackage{color}
\usepackage{soul}
\sethlcolor{yellow}
\usepackage{float}

\begin{document}

\title{\bf Anisotropic Local Correlations and Dynamics in a Relaxor Ferroelectric}

\author{ Hiroyuki Takenaka,  Ilya Grinberg, and Andrew M. Rappe}

\affiliation{The Makineni Theoretical Laboratories, Dept. of Chemistry, University of Pennsylvania, Philadelphia, PA 19104-6323, 
}

\date{\today}

\begin{abstract}
Relaxor ferroelectrics have been a focus of intense attention due to
their anomalous dielectric characteristics, diffuse phase
transitions, and strong piezoelectricity. Understanding the structure and dynamics of relaxors has been one of the long-standing challenges in solid-state physics, with the current model of polar nanoregions in a non-polar matrix providing only a qualitative description of the relaxor phase transitions.
In this paper, we investigate the local structure and dynamics in
75\%PbMg$_{1/3}$Nb$_{2/3}$O$_3$-25\%PbTiO$_3$ (PMN-PT) using molecular dynamics simulations and the dynamic pair distribution
function technique. We show for the first time that relaxor transitions can be described by local order parameters. We find that structurally, the relaxor phase is characterized by the presence of highly anisotropic correlations between the local cation displacements. These correlations resemble the hydrogen bond network in water.   Our  findings contradict the current polar nanoregion model; instead, we suggest a new model of a homogeneous random network of anisotropically coupled dipoles.
\end{abstract}

\maketitle

Recently, relaxor ferroelectrics have become important in technological applications, resulting in a revival of interest in this longstanding fundamental scientific problem.
~\cite{Park97p1804,Mischenko06p242912,Kutnjak06p956,Gehring00p5216,Blinc03p247601,Scott07p954}
Compared with normal ferroelectrics, relaxors exhibit a stronger piezoelectric effect, a high permittivity over a broad temperature range, and unique dielectric response with strong frequency dispersion.
The inverse dielectric response starts deviating from the Curie-Weiss law at the Burns temperature ($T_b$), significantly above the Curie temperature ($T_c$).
For the past several decades, these effects have been ascribed to the appearance of polar nanoregions (PNRs) which form spherical or elliptic clusters in non-polar matrix at $T_b$ due to random fields in the material, with the size and the interactions of the PNR increasing  as $T$ is lowered to the Vogel-Fulcher freezing temperature ($T_f$).   However, this model provides only a qualitative description of the changes in the structure through relaxor transitions.  Furthermore, recent studies using Raman, NMR, neutron scattering pair distribution functions (PDFs), and diffuse scattering techniques~\cite{Svitelskiy03p104107,Toulouse05p184106,Blinc99p424,Dmowski08p137602,Gehring09p224109} have demonstrated that static local  polarization on at least a nanosecond time scale appears only at a  temperature $T^*$ roughly halfway between $T_b$ and $T_f$.  

In this work, we use molecular dynamics simulations and analysis of PDFs for 75\%Pb(Mg$_{1/3}$Nb$_{2/3}$)O$_3$-25\%PbTiO$_3$ (PMN-PT)
to show that relaxor transitions are characterized by well-defined and observable local order parameters and are due to the onset of anisotropic nanoscale correlations of the in-phase cation motions. These correlations do not form clusters and therefore  cannot be explained by the current PNR  model. Rather, the couplings between displacements are analogous to the network of hydrogen bonds in water.
Both in water and in relaxors, the presence of such a network in a polar environment leads to unique physical properties.
We therefore propose that the  PNR model should be replaced with a model of a hydrogen-bond-like network of dipoles generated by local anisotropically coupled cation displacements.

We study an 8640-atom supercell of PMN-PT with bond-valence (BV) molecular dynamics (MD) simulations~\cite{Shin05p054104,Grinberg09p197601} in this work. 
Application of the dynamic pair distribution function (DPDF) method~\cite{Dmowski08p137602} enables identification of the size and directions of displacement correlations and the atomic-scale local order parameters for relaxor transitions
 (Computational details of MD and methodology of DPDF are described in supplementary information).
\begin{figure}
{
\includegraphics [width=3.5in]{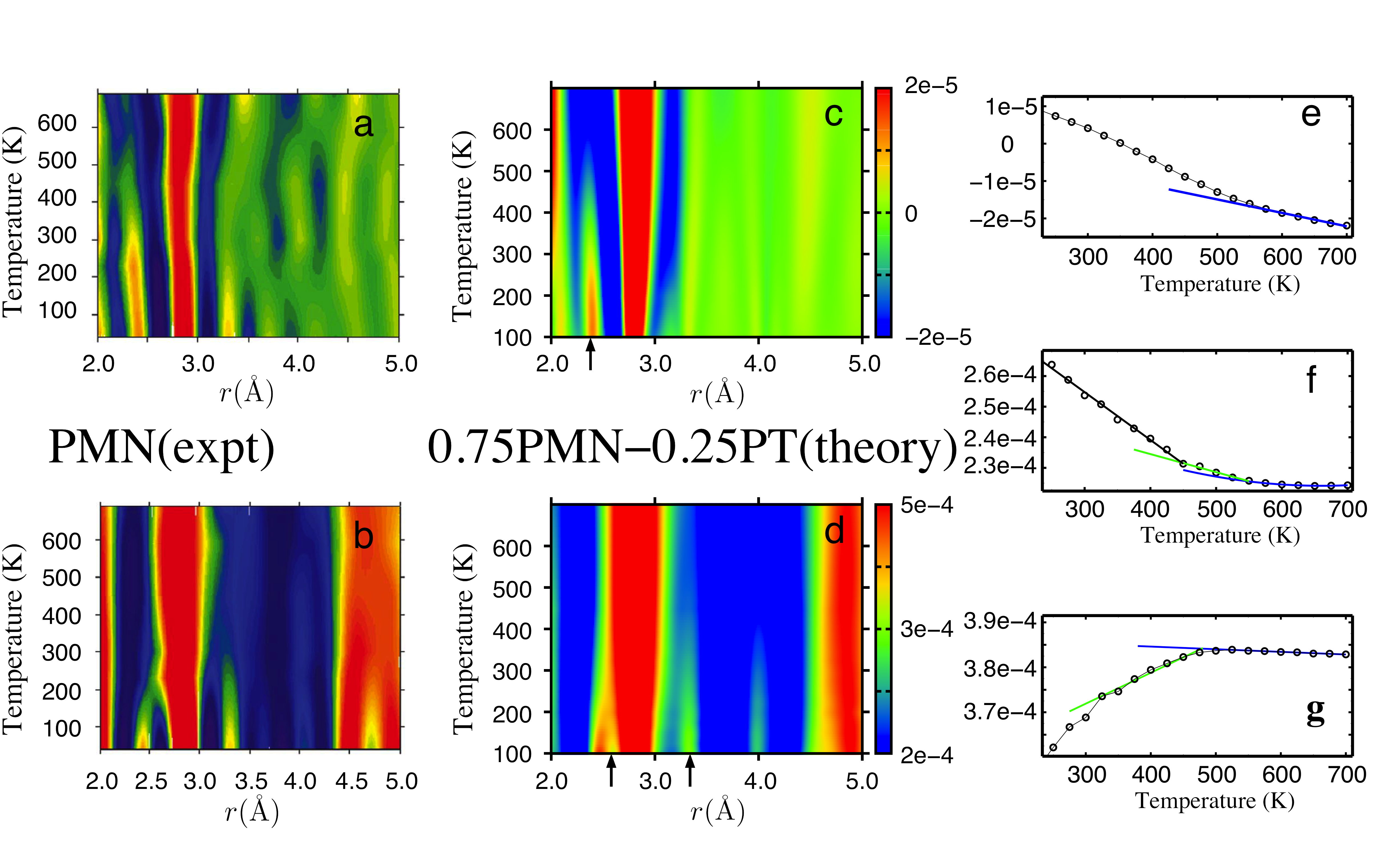}
}
\caption{(Color online) The frequency averaged DPDF.
(a-b) Color contour plot of experimental $G(r,\omega)$ for PMN integrated from 10~meV to 20~meV and from -5~meV  to 5~meV.~\cite{Dmowski08p137602}   (c-d) Color contour plot of computational 0.75PMN-0.25PT  $G(r,\omega)$  integrated from 10~meV to 20~meV and from 0 to 5~meV, respectively. Arrows mark the $r$ values for which a 1D plot of DPDF intensity versus temperature is presented in (e) integrated $G(r,\omega)$ for the 10-20~meV frequency range at $r$=2.38~\AA, (f) for the 0-5~meV frequency range at $r$=3.3~\AA, and (g) for the 0-5~meV frequency range at  $r$=2.58~\AA.
See text for interpretation.
}
\label{intdpdf}
\end{figure}

We first compare our results with the experimental 0.75PMN-0.25PT phase transition temperatures and with integrated experimental DPDF data of Dmowski {\em et al.}~\cite{Dmowski08p137602} for PMN(Figure 1a-b).
We integrate the total DPDF from  10~meV to 20~meV and from 0~meV to 5~meV for $2<r <5$~\AA\ for our 0.75PMN-0.25PT material (Figure 1c-d).  
For the higher frequency interval (Figure 1c), a short Pb-O peak appears at $\approx$550~K, same as the experimental $T_b$=550~K of 0.75PMN-0.25PT.~\cite{Dkhil09p064103}
This can be clearly seen in a 1D plot of the integrated high-frequency DPDF intensity at $r$=2.38~\AA\ versus temperature (Figure 1e), where a more rapid rise of DPDF intensity with lower $T$ sets in at 550~K.  
Though the color change in Figure 1d is subtle, the 1D plot for the integrated lower-frequency DPDF intensity at $r$=3.3~\AA\ (Figure 1f) shows two changes of slope at 550~K and at 425~K.
The transition at 425~K is at a temperature that is intermediate between the experimental $T_b$=550~K and $T_f$=380~K of 0.75PMN-0.25PT.~\cite{Bokov00p1888} 
The integrated lower frequency DPDF intensity for the short Pb-O distance (Figure 1d) shows a split of the Pb-O peak away from the main Pb-O and O-O distances peak at 2.83~\AA.
Since the splitting gives rise to a decrease of the DPDF intensity at the distance between the two peaks, we examine the integrated lower frequency DPDF intensity for $r=2.58$~\AA\ (Figure 1g).
This shows two changes of slope at 525~K and at 350~K, with the rapid intensity decrease and strong peak splitting appearing at temperatures slightly below the experimental $T_f$=380~K of 0.75PMN-0.25PT.
Similar features are present in the integrated DPDF intensities for PMN obtained by Dmowski {\em et al.} (Figure 1a-b).
Here, in integrated DPDF in the higher frequency interval (Figure 1a), the 2.4~\AA\ peak appears at $\approx$600~K, close to $T_b$=630~K of PMN. In the lower frequency (Figure1b), at $r=3.3$~\AA\ there is a strong intensity enhancement at $\approx$300~K, between PMN $T_b$=630~K and $T_f$=200~K,~\cite{Dmowski08p137602} and a faint increase in intensity above 300~K.
The 2.4~\AA\ peak visibly splits off from the main 2.8~\AA\ peak at $\approx$170~K, slightly below $T_f$=200~K.
The correspondence between the transition temperatures
found in our DPDF data and the experimental 0.75PMN-0.25PT values as well as the
good agreement between the main features of the experimental PMN and computational
0.75PMN-0.25PT DPDFs show that our simulations are a good basis for investigating the
structure and dynamics in relaxor ferroelectrics.

Cation-oxygen instantaneous and time-averaged PDFs (Figure 2) reveal the changes in the dynamics as the temperature is lowered and the system undergoes a sequence of transitions.
The  peak of the instantaneous Pb-O partial PDF $g(r,t=0)$ presented in Figure 2a corresponds to the Pb-O$_{12}$ cage and is asymmetric for all temperatures with the positively skewed distributions of $g(r,t=0)$ above $T=550$~K.
This means that Pb atoms shift away from the center of their O$_{12}$ cages even in the high-temperature paraelectric phase, in agreement with previous results.~\cite{Sepliarsky11p435902,Shin05p054104,Pasciak12p224109}
Unlike $g(r,t=0)$, the time-averaged Pb-O partial PDF $G(r,\omega=0)$ in Figure 2b shows almost symmetric Pb-O peaks centered at 2.81~\AA\ for $T>$475~K, indicating a small and temperature independent time-averaged Pb atom off-centering. Such  small time-averaged local cation displacements were observed in previous simulations and are due to the strong local random fields created by the local variations in the $B$-cation arrangement.~\cite{Sepliarsky11p435902,Grinberg09p197601}

To clearly show the changes of local structure with temperature, we use the short Pb-O peak positions for $g(r,t=0)$ and $G(r,\omega=0)$ to calculate the magnitude of the instantaneous and the time-averaged static local Pb displacements ($D_{\rm Pb}^{\rm inst}$ and $D_{\rm Pb}^{\rm static}$) shown in Figure 2c.
These are obtained by subtracting the peak position from 2.83~\AA\ Pb-O distance of the high symmetry Pb-O$_{12}$ cage at the 0.75PMN-0.25PT lattice constants.
We also show the dynamic component of the Pb displacement $D_{\rm Pb}^{\rm dyn}$ defined as the difference between $D_{\rm Pb}^{\rm inst}$ and $D_{\rm Pb}^{\rm static}$.

\begin{figure}
{
\includegraphics [width=3.0in]{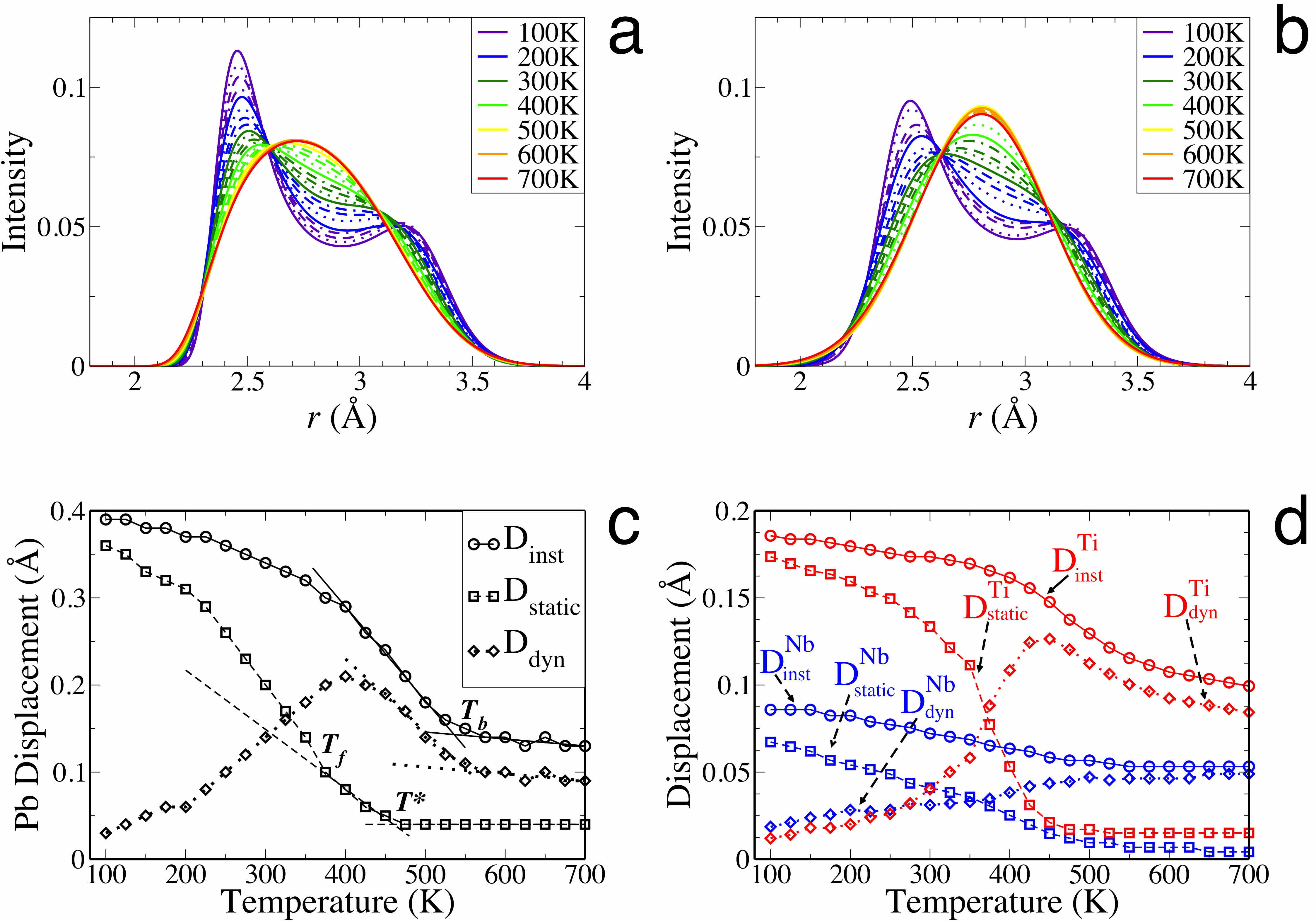}
}
\caption{(Color online) Temperature dependence of instantaneous and time averaged PDFs for cation-O atomic pairs.
(a) Instantaneous $g$($r$,$t$=0) and (b) time-averaged $G$($r$,$\omega$=0) Pb-O partial PDFs for the first Pb-O peak.
(c) The magnitude of the instantaneous Pb off-center displacement and its static and dynamic components as a function of temperature.
(d) Nb and Ti displacements as a function of temperature as obtained from the Nb-O and Ti-O $g$($r$,$t$=0) and $G$($r$,$\omega$=0).
See text for interpretation.
}
\end{figure}

Examination of the data in Figure 2c shows that three phase transitions occur at $T_b$=550~K, $T^*$=450~K and $T_f$=375~K, separating four distinct regions.
The values of all three transition temperatures are agreement with experimental 0.75PMN-0.25PT data.~\cite{Dkhil09p064103,Bokov00p1888}
For  $T>T_b$, $D_{\rm Pb}^{\rm inst}$  and $D_{\rm Pb}^{\rm static}$ are both small and change little with temperature.
For 450~K$<T<$550~K, $D_{\rm Pb}^{\rm inst}$ increases rapidly as $T$ is lowered, while $D_{\rm Pb}^{\rm static}$ is the same as for $T>550$~K. The increasing dynamic displacement component $D_{\rm Pb}^{\rm dyn}$ indicates the onset of the dynamic relaxor phase.
For 400~K$<T<$450~K, $D_{\rm Pb}^{\rm static}$ increase slowly and  $D_{\rm Pb}^{\rm dyn}$ starts to plateau.   Finally, at $T$=375~K, another transition takes place with  $D_{\rm Pb}^{\rm static}$ increasing rapidly and $D_{\rm Pb}^{\rm dyn}$ decreasing and both then saturating at their low temperature values, as the system undergoes a transition into the frozen phase.

The transition at $T^*$ is characterized by the change in $D_{\rm Pb}^{\rm static}$; we therefore assign the difference between the calculated $D_{\rm Pb}^{\rm static}$ and the $D_{\rm Pb}^{\rm static}$ of the paraelectric phase  as the order parameter for the low-temperature phases at $T<$475~K.
The $D_{\rm Pb}^{\rm static }$  is related to the Edwards-Anderson spin glass order parameter $q$,  defined as $q$=$\Sigma_i <S_i>^2$~\cite{Edwards75p965}, which measures the average magnitude of the static local polarization in the material and has been used to model relaxor behavior.~\cite{Blinc99p424,Sepliarsky11p435902,Akbarzadeh12p257601}
Using NMR experiments on PMN  and an analytical random-bond random-field spin model, Blinc {\em et al.} have previously shown that $q$ rises above zero at a temperature between $T_b$ and $T_f$.~\cite{Blinc99p424}
This is similar to our finding of  $D_{\rm Pb}^{\rm static }$ rising above the paraelectric value at $T^*$=450~K. 
Since  $q$ and $D_{\rm Pb}^{\rm static}$ {\em do not} change at $T_b$, the freezing of the local displacements that has been the focus of previous investigations is not the relevant process for the onset of relaxor behavior at $T_b$.
Rather, the transition at $T_b$ is characterized by the increase in the local scalar magnitude of the instantaneous cation displacements $D_{\rm Pb}^{\rm inst}$  (Figure 2c).
We therefore assign the difference between the calculated $D_{\rm Pb}^{\rm inst}$ and the $D_{\rm Pb}^{\rm inst}$ of the paraelectric phase  as the order parameter for the dynamic relaxor phase found for $\approx$475~K$<T<$550~K.  

Inspections of Nb-O and Ti-O partial PDFs (Figure 2d) show trends similar to those observed for Pb displacements (Mg cation displacement magnitudes are close to zero for all temperatures).   The main difference between the Pb and $B$-cation displacements is that the dynamic component of the off-center displacement $D^{\rm dyn}$ peaks at a higher temperature for the $B$-cations than for the Pb atoms. The cation-oxygen bonding is more distributed and flexible in the PbO$_{12}$ cuboctahedron than in the $B$O$_6$ octahedron. This makes the freezing temperature of the Pb displacements lower than for the Nb and Ti cations. 

The fact that a local quantity such as Pb displacement magnitude shows order parameter behavior implies that local interactions are at the root of relaxor behavior.
Enhanced local correlation between the Pb displacements is the physical origin of the increase in $D_{\rm Pb}^{\rm inst}$ for $T<550$~K.
Pb atom displacements in neighboring unit cells are  coupled, so that correlated displacements in the same direction (even if time-averaged to zero) allow a greater magnitude of individual Pb ion off-centering.   
We now examine the cation-cation DPDFs to reveal the changes in displacement-displacement correlations that give rise to the structural and dynamical properties of the relaxor phase.

\begin{figure}
{
\includegraphics [width=3.3in]{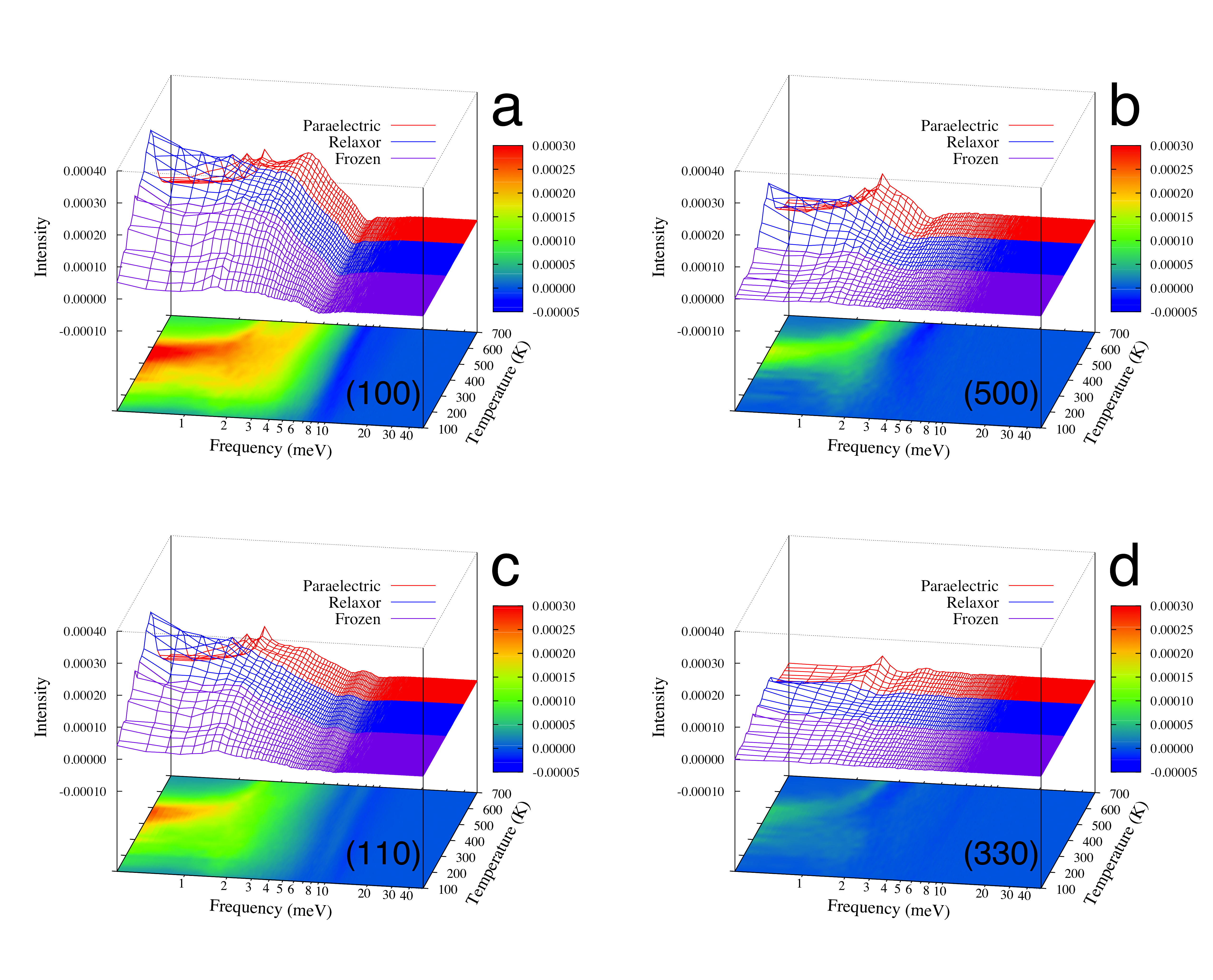}
}
\caption{(Color online)
Pb-Pb $G$($r,\omega$) along the high-symmetry directions  as functions of temperature (K) and frequency (meV) at the peak position of the instantaneous PDF for each temperature. The  contour is projected at bottom with a color bar which shows the intensity of the peaks.  Data for (a) $r$=(100) (b) $r$=(500) (c)$r$=(110) (d)$ r$=(330) are shown.
}
\end{figure}

We compare  Pb-Pb $G(r,\omega)$ along the (100), (110) and (111) high symmetry directions.
In Fig.3, we show only the (100) and (110) directions since the intensities along (111) directions are only slightly weaker than the intensities along (110) directions.
The $G(r,\omega)$ reveal the spatial extent and the frequency spectrum of the correlated cation motion. 
We find that  the correlations between Pb displacements are enhanced between $T_f$ and $T_b$ and exhibit a strong direction dependence.
Starting at $T_b$, the $G(r,\omega)$ data along the (100) direction show a strong increase in intensity at low $\omega$, with a shift of the low frequency peak to below 0.1 meV  as $T$ approaches $T_f$.
The intensity weakens and the peak position changes as $r$ increases from 4~\AA\ to 20~\AA, but the appearance of low-$\omega$ peaks at $T_b$ and their shift to lower frequencies with lower $T$ is present for all DPDF along the (100) direction.
By contrast, although the (110) and (111) directions also show peaks in the in-phase vibration intensity, they decay dramatically with increasing distance. For example, for all temperatures, we find essentially zero DPDF intensity for the (330) Pb-Pb peaks, which indicates  a lack of correlated in-phase oscillations.
This shows that below $T_b$, the coupling between the local dipoles created by Pb displacements is anisotropic, with strong interactions only between the dipoles located  along Cartesian directions.  
Anisotropic correlations between cation displacements were also very recently reported by Akbarzadeh et al. in Ba(Zr,Ti)O$_3$ relaxor.~\cite{Akbarzadeh12p257601}
The direction dependences of the $B$-cation-$B$-cation correlations are more
substantial and anisotropically strong coupling, confined to $B$-cation sublattice, along the
Cartesian directions is present even in the high temperature paraelectric phase of PMN-PT.
 (See in supplementary information).
This is in disagreement with the current model of PNR inside a non-polar matrix, where  correlations should extend along all directions as the temperature is lowered below $T_b$.

The stronger correlation along the (100) direction indicates that the Pb displacement coupling is not solely due to the dipole-dipole interactions.
The anisotropy is induced by the through-oxygen interactions between nearest-neighbor cations that share one (for the $B$-cations) or more  (for Pb) O atoms along a Cartesian direction in the $AB$O$_3$ structure.
These interactions are stronger for the $B$-cations, due to the much higher average bond valence of each $B$-cation-oxygen bond (4/6) compared to the valence of the average Pb-O bond (1/6).
The sharing brings strong $B$-cation displacement coupling even in the paraelectric phase.
On the other hand, for Pb atoms, the coupling along (100) is weaker and is therefore absent above $T_b$.  

Despite the presence of strong $B$-cation-$B$-cation displacement correlations, at $T>T_b$ PMN-PT still exhibits normal paraelectric behavior.
We ascribe this to the fact that the strongly correlated $B$-cation displacement chains are one dimensional and therefore cannot undergo a phase transition into a more ordered phase.
In addition to the through-oxygen interactions, the chains of the $B$-cation-$B$-cation correlations enhance Pb atom correlations along the Cartesian directions.
At $T_b$, as the Pb atom displacements next to the strongly correlated $B$-cation chains become correlated, the correlated region becomes three-dimensional, enabling a phase transition from a disordered paraelectric phase to a locally correlated dynamic relaxor phase.

Our results show that the region of correlation is strongly anisotropic, with a network of tube-like correlations along the Cartesian axes and a radius of around 7.5~\AA\ in the other directions.
This is the same as $\approx7$~\AA\ estimated for the size of PNR by Gehring {\em et al.} based on the "waterfall effect" in 0.8PMN-0.20PT.~\cite{Gehring00p5216}
The change of the low $\omega$ part of the spectrum from a sharp peak at $T>T_b$ to a broad band absorption for $T< T_b$ is in agreement with the waterfall effect observed by neutron-scattering experiments in relaxor materials.
A waterfall dispersion relation will result in a density of states with a peak at higher $\omega$ and a flat region on the low-frequency side of the peak, precisely as observed in our $G(r,\omega$) for $T<T_b$.
Therefore, our simulations suggest that the experimentally observed waterfall effect is due to the correlated in-phase vibrations of cation pairs coupled by short-range through-oxygen interactions. 

The correlations between cation displacements show a strong dependence on the identity of the atoms.
For low $\omega$, high intensity of in-phase vibrational correlation is favored by Nb and is disfavored by Mg, with Ti intermediate.
This is particularly pronounced for the Pb-$B$-cation $G(r,\omega)$  where the Pb-Mg $G(r,\omega)$ shows weaker changes with temperature for the short $r$=3.5~\AA\ (111) peak (See in supplementary information), in contrast to all other atom pairs at $r<4$\AA\ (including Mg-Mg).
Previous BVMD simulations have found that Mg atoms and Nb atoms surrounded by Mg atoms (Nb$^{\rm MM}$) exhibit fast Arrhenius dynamics typical of the high-temperature paraelectric phase even for $T< T_f$.~\cite{Grinberg09p197601,Grinberg12preprint,Kamba07p074106}
Since Mg-Mg, Mg-Nb and Mg-Ti $G(r,\omega)$ all exhibit the intensity at low $\omega$ associated with relaxor behavior, we assign the lack of Pb-Mg coupling as the cause of the paraelectric-like dynamics of the Mg and Nb$^{\rm MM}$ atoms.  

The current understanding of the relaxor transitions is that at $T_b$ small and dynamic spherical PNR form within a paraelectric matrix; as $T$ is lowered, the PNR grow, show smaller polarization fluctuations, and  start to freeze in at $T^*$.~\cite{Cowley11p229,Pirc09p254,Gehring09p224109}  At $T_f$, a percolation transition takes place as all of the PNR coalesce into a a single cluster resulting in the freezing of the local polarizations.~\cite{Pirc07p020101}  The static relaxor state is described as inhomogeneous and consisting  of static PNR in a statically non-polar matrix, with volume fraction of the PNR of about 0.3.~\cite{Jeong05p147602}  

Our results cast doubts on the conventional picture of relaxor structure and dynamics. Recent MD simulations have shown that structural and dielectric properties of PMN-PT can be reproduced without PNR.~\cite{Grinberg09p197601,Sepliarsky11p435902}  Relaxor behavior in Pb-based perovskites was also shown to be controlled by local structure parameters.~\cite{Grinberg07p267603}
In this work, we find that the average magnitude of local cation displacements is quite large at  $T\le T_f$=375~K, with $D_{\rm Pb}$ magnitude of $\approx$0.40~\AA\!, close to that found in the prototypical normal ferroelectric PbTiO$_3$.
In ferroelectric perovskites, uncorrelated displacements incur a large energy cost due to  oxygen atom under- and overbonding, increased $A$-$B$ cation repulsion, and unfavorable dipole-dipole interactions.~\cite{Grinberg04p220101,Grinberg07p037603} 
Therefore, a large magnitude of local polarization requires strong correlated displacements, as found in our simulations of PMN-PT; this is inconsistent with the idea that most of the material exists in a non-polar matrix state.   Additionally, we find that correlations between the cation displacements are highly anisotropic and are weak for (110) and (111) directions even at temperatures close to $T_f$; this is inconsistent with the picture of the strongly polarized spherical nanoclusters.  

We therefore suggest a alternate model for relaxor structure and dynamics.
The transition at $T_b$ is characterized by a shift from dipole-dipole interactions to the anisotropic short-range, through-oxygen coupling along the Cartesian directions.
At high $T$, the thermal energy is high enough to disrupt the through-oxygen coupling, so that dipole-dipole interactions play a dominant role in determining the dynamics of the system.  
At $T_b$, the energy lowering due to correlated displacements and the decrease in oxygen atom overbonding is larger than the entropy cost of correlated displacements.
Therefore, the structure changes to a network of strong, through-oxygen coupled motions along the Cartesian axes.  
Such a transition is similar to the changes that take place when superheated water is cooled down to room temperature, where a standard polar liquid local structure dominated by dipole-dipole interactions is transformed into a directional, highly anisotropic random H-bond network with unique structural and dynamical properties.
Extending an analogy suggested by Pirc and Blinc relating the ferroelectric phase to solids and the paraelectric phase to liquids,~\cite{Pirc09p254} we suggest that relaxors are analogous to hydrogen-bonded water.
A coupling network  mediated by O atoms plays  the role of the H-bonds in water and preferential bonding directions lead to dynamic clusters of correlated displacements of various sizes.
The fact that water exhibits Vogel-Fulcher dielectric response including some extremely slow relaxation processes due to the collective H-bond network behavior further supports this analogy.~\cite{Jansson10p017802}

I.G. and A.M.R were supported by the Office of Naval Research, under Grant No. N00014-11-1-0578. H.T. was supported by the NSF under grant DMR11-20901.
Computational support was provided by a Challenge Grant from the HPCMO of the U.S. Department of Defense.

\newpage
\section*{\Large{Supplementary information}}
\renewcommand{\thesection}{\bf S\!\! \arabic{section}}
\renewcommand{\tablename}{\bf Table S\!\!}
\renewcommand{\thetable}{\arabic{table}}
\renewcommand{\figurename}{\bf Fig. S\!\!}
\renewcommand{\thefigure}{\arabic{figure}}
\setcounter{equation}{0}
\setcounter{figure}{0} 
\section{Computational details}
We use a 12$\times$12$\times$12 0.75P(Mg$_{(1/3)}$Nb$_{2/3}$)-0.25PbTiO$_3$(PMN-PT) supercell with random-site ordered B-cation arrangement~\cite{Akbas97p2933} and perform MD simulations at a range of temperatures from 100~K to 700~K using an atomistic bond-valence potential~\cite{Shin05p054104} derived from first-principles calculations and used successfully in a previous study of 0.75PMN-0.25PT.~\cite{Grinberg09p197601} The BV potential parameters in this work are shown in Table S1 and S2,  with the O$_6$ tilt angle potential parameter set to $1.6875 \rm{meV}/(\rm {deg})^2$.  We run our simulations for 0.5~ns in order to be able to extract low frequency data.
\begin{table}[H]
\begin{center}
\begin{tabular}{ccccc}
\hline   & $q_\beta$ (e)  & $S_\beta$ (eV)  & $r_0^{\beta\beta^{\prime}}$ & $C_{\beta\beta^{\prime}}$   \\
\hline
Pb &    1.574       & 0.0618  & 2.044 & 5.50   \\
Mg &    0.185       & 1.1555  & 1.622 & 4.29   \\
Nb &    0.503       & 0.3874  & 1.907 & 5.00   \\
Ti &    0.575       & 0.5839  & 1.806 & 5.20   \\
O  &   -0.672       & 0.4976  & N/A    & N/A   \\
\hline
\end{tabular}
\caption{\rm{The charges, $q_{\beta}$, and bond-valence model parameters, $S_\beta$, for the atomistic model of 0.75PMN-0.25PT. The values of $r_0^{\beta\beta^{\prime}}$ and $C_{\beta\beta^{\prime}}$ we
re empirically determined by Brown.\!\!\!~\cite{Brownbook} The notations accord with Reference2.}}
\end{center}
\label{table:supercell-table}\end{table}
\vspace{-0.1in}
\begin{table}[H]
\begin{center}
\begin{tabular}{cccccc}
\hline
   &   Pb     & Mg    & Nb    & Ti    & O    \\
\hline
Pb &    2.221 & 0.553 & 2.454 & 2.312 & 1.676 \\
Mg &    0.553 & 1.344 & 2.920 & 1.442 & 1.415 \\
Nb &    2.454 & 2.920 & 0.408 & 3.053 & 1.278 \\
Ti &    2.312 & 1.442 & 3.053 & 1.037 & 1.181 \\
O  &    1.676 & 1.415 & 1.278 & 1.181 & 1.927 \\
\hline
\end{tabular}
\caption{\rm{The parametrized values of coefficients in the interatomic repulsive potentials, $B_{\beta\beta^{\prime}}$ in Reference 2, for our 0.75PMN-0.25PT model. }}
\label{table:supercell-table}
\end{center}
\end{table}

\section{Dynamic Pair Distribution Function Analysis Methodology}
In this section, we describe procedure of dynamic pair-distribution function(DPDF) calculation and explain its physical meanings and usefulness.
Using MD simulation results, we then calculate the $g$($r$,$t$) generalized PDF which measures the probability of the position of one atom at time $t'$ being separated by a distance
 $r$ from the position of the other atom at time $t''$, such that $t'-t''$=$t$. The $g(r,t)$ denotes the correlation function at a given distance $r$ and time delay $t$:
\begin{eqnarray}
g(r,t)&=&
\frac{1}{N\langle b\rangle^2}
\sum_{\nu,\mu=1}^N
\frac{1}{\sqrt{\pi \sigma_{\nu}+\sigma_{\mu}}}
b_{\nu}b_{\mu}
\frac{1}{4\pi r^2} \times\cr
& &\int
\!\!e^{-\frac{(r-|R_{\nu}(t^{\prime})-R_{\mu}(t+t^{\prime})|)^2}{\sigma_{\nu}+\sigma_{\mu}}}dt^{\prime},
\label{eqtdpdf}
\end{eqnarray}
where $N$ is the number of atoms, $b$ is a neutron scattering length,$\langle b \rangle$ is the average of the scattering lengths, $\sigma$ is the normalization factor of the Gaussian smearing width, and $R$ is an atomic position.
The $g(r,t)$ can be Fourier transformed to $G(r,\omega)$ which denotes DPDF.  This makes it possible to resolve the vibrational frequencies for the different types of atomic pairs at different distances, relating $G(r,\omega)$ results to the experimental studies of dynamics. We use three different PDFs in this work for structural and dynamics analysis:  $g$($r$,$t$=0), $G$($r$,$\omega=0$) and $G$($r$,$\omega$). The integral of $G$($r$,$\omega$) over $\omega \neq 0$ gives the difference between $g$($r$,$t$=0) and $G$($r$,$\omega=0$). The physical interpretation of these different PDF types can be illustrated as follows.
Consider a generalized PDF $g_{AB}(r,t)$ defined as the probability of an atom of species $A$ at time $t'$ located at distance $r$ from the position of an atom of species $B$ at time $t''$, such that $t'-t''$=$t$.   Then $g_{AB}(r,t=0)$ is the instantaneous PDF.  A peak in $g_{AB}(r,t=0)$ is broadened due to structural variation in the preferred interatomic distance $r$ and due to out-of-phase vibrations of the $A-B$ pairs around the minimum energy $r$.  The in-phase vibrations preserve the preferred distance $r$ and do not broaden the peak. The widely used elastic neutron scattering technique measures the instantaneous PDF.  Similar to the peaks in $g_{AB}(r,t=0)$, a peak in $G_{AB}$ is broadened by structural variation in the preferred interatomic distance $r$ and by out-of-phase vibrations of the $A-B$ pairs around the minimum energy $r$.  Additionally, it is also  broadened by the in-phase vibrations of $A-B$ pairs.  This means that a difference 
between $g_{AB}(r,t=0)$  and $G_{AB}(r,\omega=0)$ specifies the amplitude and the intensity of the in-phase vibrations of the ions.   The total difference $g-G$ can be decomposed into contributions at each $\omega$.  Thus, DPDF analysis identifies at which frequencies particular types of atom pairs at particular distances are vibrating in-phase.  This capability is particularly relevant to the study of relaxor dynamics where the appearance of in-phase vibrations of the cations characterizes the transition from the paraelectric to the relaxor phase.

\section{Pb-Pb DPDF along the (111) direction}
\begin{figure}[H]
{\includegraphics [width=3.30in]{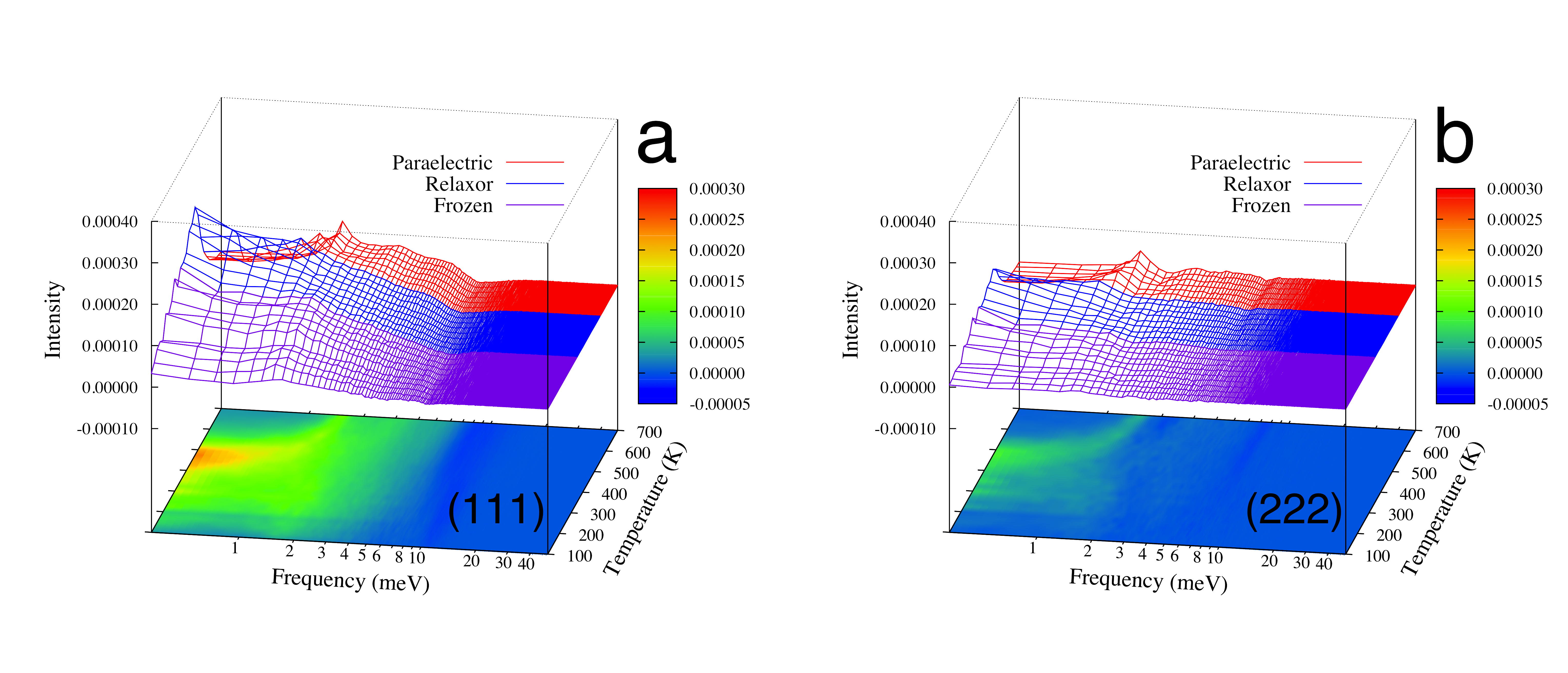}
}\caption{\rm{(Color online) First and second neighbor Pb-Pb DPDF along (111) directions as functions of temperature (K) and frequency (meV) at the peak height of the instantaneous PDF for each temperature. The contour is projected at bottom and a color bar the intensity of the peaks.  Data for (a) $r$=(111) (b) $r$=(222)
Correlations of in-phase low-frequency vibration along the (110) and the (111) directions at $r> 8$~\AA\ get weaker even than the correlations at $r$=(500).}
}
\label{pbpbdpdf}
\end{figure}

The DPDF, $G(r,\omega)$, of Pb-Pb atomic pairs for $r$=(111) and $r$=(222) are shown in  Fig.S\ref{pbpbdpdf}.  Similar to the DPDFs along the (110) direction, there is a rapid decay in peak intensity  with distance.

\section{Nb-Nb DPDF}
\begin{figure}[H]
{
\includegraphics [width=3.30in]{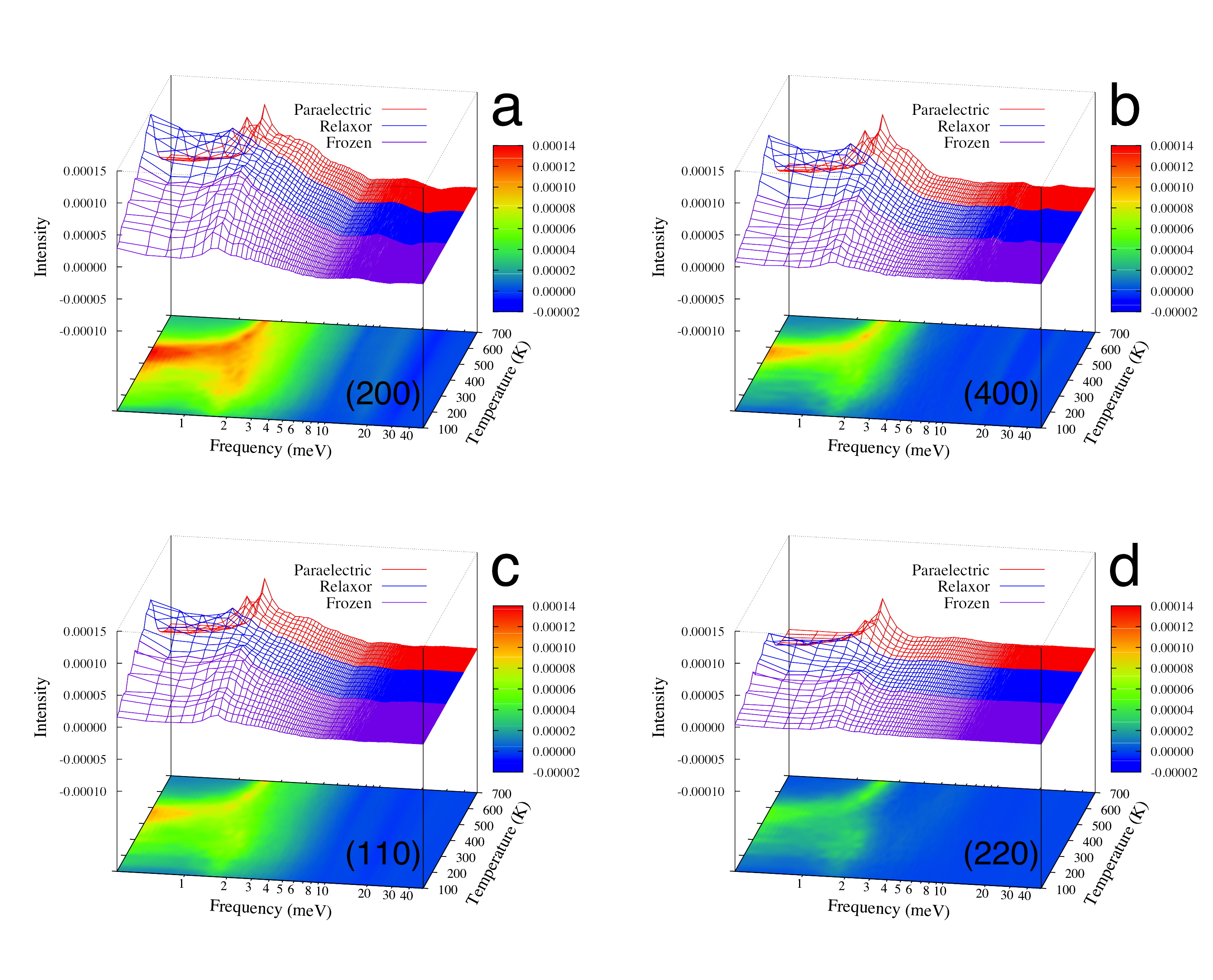}
}
\caption{\rm{(Color online) Nb-Nb DPDF along the high-symmetry directions  as functions of temperature (K) and frequency (meV). Distance corresponds to the peak position of the instantaneous PDF for each temperature. The contour is projected at bottom with a color bar which shows the intensity of the peaks.  Data for (a) $r$=(200) (b) $ r$=(400) (c)$r$=(110) (d)$ r$=(220) are shown. 
For the (100) direction,  enhanced intensity at low $\omega$ for  $T<T_b$ is present even for large distances. For the (110) and the (111) directions low-frequency vibrations disappear for $r> 8$~\AA.   In-phase vibrational intensity at higher $\omega$ is more intense for the Cartesian peaks than for non-Cartesian peaks for all $T$.}
}
\label{nbnbdpdf}
\end{figure}

DPDF peaks, $G(r,\omega)$, for Nb-Nb atomic pairs are shown in Fig.S\ref{nbnbdpdf}.
The $G(r,\omega)$ for the (400) Nb-Nb peak shows strong positive intensity, whereas $G(r,\omega)$ for the (220) and the (222) Nb-Nb peaks show almost no positive intensity despite the shorter interatomic distance. However, here the intensity for the ($n$00) peaks is quite strong already in the paraelectric phase.  This is different from the behavior observed for the Pb atoms in Figure 3 of main text, where in the paraelectric phase the in-phase vibrations are only weakly enhanced for the ($n$00) peaks compared to the ($nn$0) and the ($nnn$) peaks.

\section{Pb-Mg DPDF}
\begin{figure}[h]
{
\includegraphics [width=2.00in]{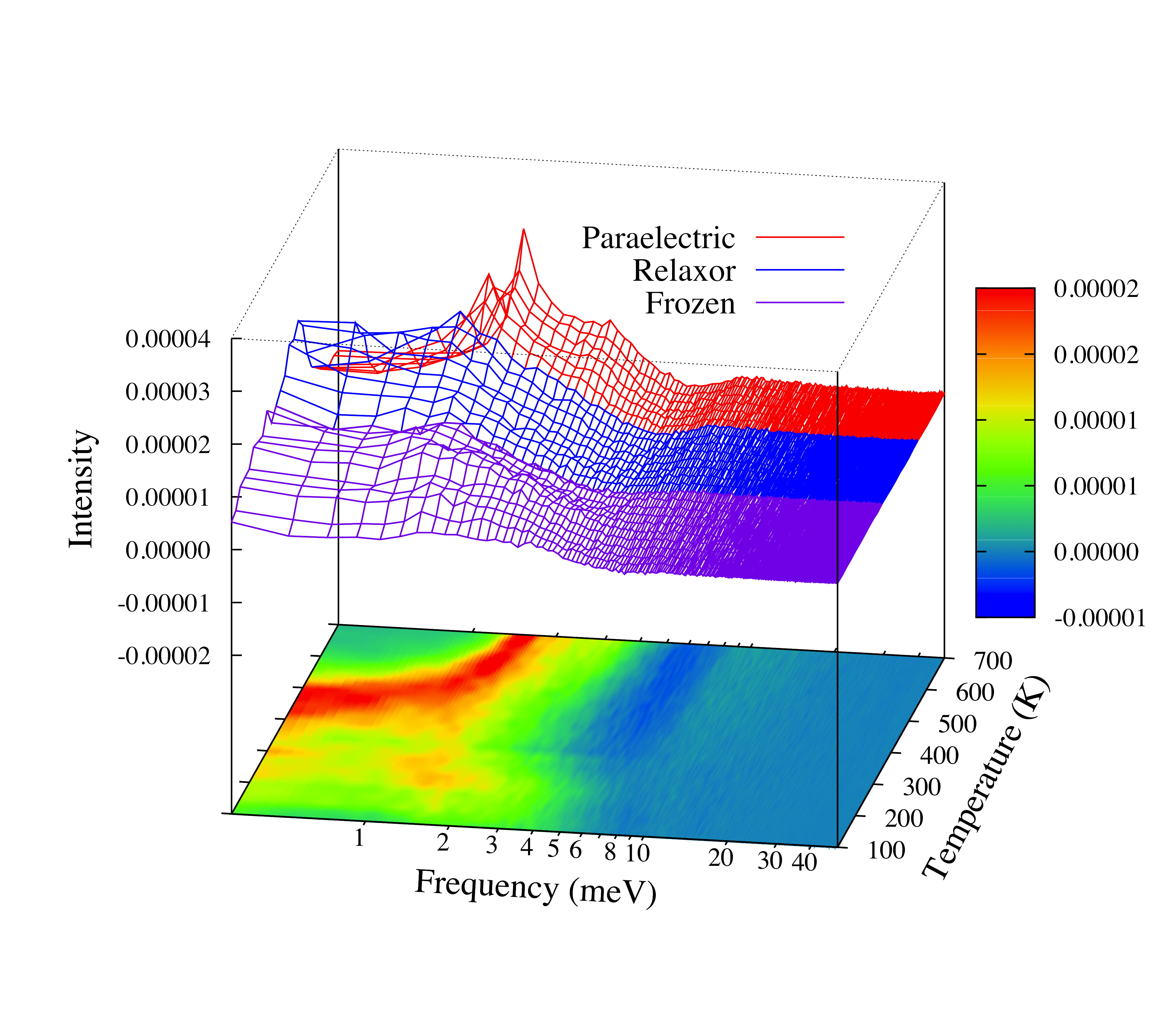}
}
\caption{\rm{(Color online) Pb-Mg DPDF peaks as function of temperature (K) and frequency (meV). The intensity values taken at the peak position of the first Pb-Mg peak in the instantaneous PDF ($r\sim 3.5$~\AA) at each temperature. The corresponding contour is projected at bottom with a color bar which shows intensity of the peaks.  Despite a shorter distance, the  enhancement of intensity at low $\omega$ for $T<T_b$ is weaker than for Pb-Pb $G$($r,\omega$) at $r=4$~\AA.}
}
\label{pbmgdpdf}
\end{figure}

The $G(r,\omega)$ peaks for $r$=(111)  Pb-Mg atomic pairs are shown in Fig.S\ref{pbmgdpdf}.  
Despite a shorter distance, the  enhancement of intensity at low $\omega$ for $T<T_b$ is weaker than for Pb-Pb $G$($r,\omega$) at $r=4$~\AA.  Such weak coupling is unique compared with other nearest-neighbor cation-cation atomic pairs in the PMN-PT.

\end{document}